\documentclass[review]{elsarticle}

\usepackage{amssymb}
\usepackage[utf8]{inputenc}
\usepackage{natbib}
\usepackage{graphicx}
\usepackage{caption}
\usepackage{subcaption}
\usepackage{tabularx}
\usepackage{color}
\usepackage{cite}
\usepackage{lineno}
\usepackage{hyperref}
\usepackage{multirow}
\usepackage[final]{pdfpages}
\usepackage{fancyvrb}
\usepackage{amsmath,amsfonts,amsthm,amssymb}
\usepackage{setspace}
\usepackage{Tabbing}
\usepackage{color}
\usepackage{fancyhdr}
\usepackage{lastpage}
\usepackage{extramarks}
\usepackage{chngpage}
\usepackage{soul,color}
\usepackage{graphicx,float,wrapfig}
\usepackage{framed}
\usepackage{color}
\usepackage{epstopdf}
\usepackage{capt-of}
\usepackage{mathrsfs} 
\usepackage{graphicx}
\usepackage{braket}
\setcitestyle{square}

\journal{Chemical Physics Letters}

\begin{document}

\begin{frontmatter}


 \author{Chinmoy Samanta}
 \ead{samantachinmoy111@gmail.com}

\title{A new approach in an analytical method for diffusion dynamics for the presence of delocalized sink in a potential well: Application to different potential curves}

\address{School of Basic Sciences, Indian Institute of Technology Mandi,
Kamand, Himachal Pradesh, 175005, India }

\begin{abstract}
\noindent We provide a new approach to solve one dimension Fokker-Planck equation in the Laplace domain for the case where a particle is evolving  in a potential energy curve in the presence of general delocalized sink. We also calculate rate constants in the presence of non-localized sink on different potential energy curves. In the previous method, we need to solve matrix equations to calculate rate constant but in our method it is not required. We also calculate the rate constant by using the method to some known potential curve, viz, flat potential, linear potential and parabolic potential.

\end{abstract}


\end{frontmatter}


\section{Introduction}
\label{}
The dynamics of diffusion motion of a particle in presence of sink on the potential energy surface (PES) provide a significant contribution to understanding various phenomena in chemical and biological process. It is modeled mathematically with the use of a Smoluchowski-like equation.
This type of model has been used to calculate rates of diffusion-controlled reactions as well as cyclization of polymers chains in solutions\citep{wilemski1974diffusion,wilemski1974diffusion2}. To describe the electron transfer reactions in polar solvent many authors have used the model of such a type\citep{ovchinnikova1982frequency,zusman1983theory,sumi1986dynamical}. A model of such a kind has been utilized by Sumi to explain the pressure influence on some isomerization reactions\citep{sumi1995}. Even the one dimensional model of diffusion motion have been using to model several problems involving proton, an activated barrier crossing or barrierless reaction process such as relaxation from electronic excited state in solution \citep{bagchi1983theory,bagchi1990dynamics,sebastian1992theory,chakravarti1993theory, samanta1990deuterium}. In a biological process, the theoretical study of the phenomena including artificial photosynthetic machines, enzyme-ligand binding and home range formation in animal behavior \citep{wade1998electrostatic,livesay2003conservation,giuggioli2006theory} has often engaged the same diffusion equation. Due to absorption of light particle drifts over the excited state of PES and getting relaxed via radiative or nonradiative decay from anywhere from the surface. In most of the model the sink term often consider to be a single Dirac delta function on the excited PES. In 1984 Szabo, Lamm and Weiss \citep{szabo1984localized} provided a generalized method when the sink is a sum of $\delta$ functions and the solution can be written in terms Green's function when there is no sink. Later Samanta and Ghosh provided an extension to the previous method and explored it for a general delocalized sink \citep{samanta1993exact}. But In their technique, it's essentially needed to solve a matrix equation and moreover, the dimension of the matrix is proportional to a number of $\delta$ function sink present on the PES. So whenever we generalize the form of sink term a large number of $\delta$ functions is needed and thus the complexity of the rate calculation is therefore reflected.\\
The purpose of this paper is to construct a new method to find the solution of Smoluchowski like equation for general delocalized sink in the Laplace domain. Our method doesn't include any matrix calculation. It is quite simple and easy to calculate the rate constant. We also know that it is not easy to calculate the non-radiative rate constant in the Laplace domain and special treatment has to make for different PES in calculating the same. The beauty of our method is if we have the non-radiative decay rate expression for a single $\delta$ function sink then we can calculate the same decay rate for the rest of $\delta$ functions sink smoothly by the recursion formula, which will be shown below. Further we apply this method for different potential energy curve (PEC) and have calculated rate constant numerically which is shown graphically.

\section{Methodology}
It is quite common to assume the motion on the PES to be one-dimensional and diffusive, the relevant coordinate being denoted by x. It's additionally usual to assume that the motion on the potential energy curve (PEC) is overdamped.
The probability distribution $P(x,t)$ corresponding to diffusion in a potential well $U(x)$ in the presence of sink function $S(x)$ is governed by the Fokker-Planck equation in the Smoluchowski form which is

\begin{equation}
      \frac{\partial P(x,t)}{\partial t}=[\mathscr{L}-S(x)-k_{r}]P(x,t).
\end{equation}
In the above, $$\mathscr{L}=D(\frac{\partial^2}{\partial x^2}+ \frac{1}{k_{B}T}\frac{\partial}{\partial x}\frac{d U(x)}{d x}).$$
 $k_{r}$ is position independent radiative decay rate and D is diffusion coefficient at temperature $T$ and $k_{B}$ is the Boltzmann constant. The above equation will often be called as a Smoluchawski equation when $U(x)$ is considered to be quadratic in $x$. Our method starts with taking the Laplace transform of Eq. (1), which is  
\begin{equation}
     [s+k_{r}-\mathscr{L}+S(x)]\mathscr{P}(x,s)=P(x,0),
\end{equation}
where $s$ is Laplace variable and $\mathscr{P}(x,s)$ denotes Laplace transform of $P(x,t)$. The initial distribution is assumed to be localized at $x_{0}$ and thus $P(x,0) = \delta(x-x_{0})$. The solution of the Eq. (2) in terms of Green's function  is given by
 \begin{equation}
     \mathscr{G}(x,s+k_{r}|x_{0})=<x| [s+k_{r}-\mathscr{L}+S(x)]^{-1}|x_{0}>.
     \end{equation}
  The solution in Laplace domain will be  
\begin{equation}
    \mathscr{P}(x,s+k_{r}|x_{0})=\int_{-\infty}^{\infty}dx'\mathscr{G}(x,s+k_{r}|x')P(x',0).
\end{equation} 
By using the operator identity of quantum mechanics we can get
     \begin{equation}
         [s+k_{r}-\mathscr{L}+S(x)]^{-1}=[s+k_{r}-\mathscr{L}]^{-1}-[s+k_{r}-\mathscr{L}]^{-1}S(x)[s+k_{r}-\mathscr{L}+S(x)]^{-1}.
     \end{equation}
The forms of the potential and also the sink function rely upon the physical problem of interest. In several cases, the potential is chosen to be parabolic and also the simplest sink function is the localized sink at an acceptable location $x_{i}$ and it is given by $S(x)=k_{i}\delta(x-x_{i})$. Although in majority problem of interest the consideration of non localized sink function is required for a proper description of dynamics. We consider the sink function as a linear combination of $\delta$ function. This form of the sink function was used by Szabo, Lamm, and Weiss \citep{szabo1984localized}. Samanta and Ghosh \citep{samanta1993exact} also used the same form of the sink function. This function is quite general because an arbitrary sink function can be written as $\int_{- \infty}^{\infty} dx'S(x')\delta(x-x')$. Farther we can discretize the integral as $S=\sum_{i}S_{i}$, where $S_{i}=k_{i}\delta(x-x_{i})$ and $k_{i}$ is strength of the $i$th $\delta$ function sink. By putting the operator identity in Eq. (5) into Eq. (3) we get the Green's function for a single sink function, $S_{1}=k_{1}\delta(x-x_{1})$, which is 
\begin{equation}
\mathscr{G}_{1}(x,s+k_{r}|x_{0})=\mathscr{G}_{0}(x,s+k_{r}|x_{0})-\frac{k_{1}\mathscr{G}_{0}(x,s+k_{r}|x_{1})\mathscr{G}_{0}(x_{1},s+k_{r}|x_{0})}{1+k_{1}\mathscr{G}_{0}(x_{1},s+k_{r}|x_{1})}.
\end{equation}
In the above $\mathscr{G}_{0}(x,s+k_{r}|x_{0})$, the Green's function corresponding to propagation of the particle placed initially at $x_{0}$ when there is no sink on the potential energy curve, is defined as\\
\begin{equation}
    \mathscr{G}_{0}(x,s+k_{r}|x_{0})=<x|[s+k_{r}-\mathscr{L}]^{-1}|x_{0}>.
\end{equation}\\
Now we add another sink function, $S_{2}=k_{2}\delta(x-x_{2})$, along with the first one and the corresponding operator identity in Eq. (5) will be
\begin{equation}
    [s+k_{r}+k_{0}[S_{1}(x)+S_{2}(x)]-\mathscr{L}]^{-1}=[s+k_{r}+k_{0}S_{1}(x)-\mathscr{L}]^{-1}
\end{equation}
$$-[s+k_{r}+k_{0}S_{1}(x)-\mathscr{L}]^{-1}k_{0}S_{2}(x)[s+k_{r}+k_{0}[S_{1}(x)+S_{2}(x)]-\mathscr{L}]^{-1}.$$
Like the case of a single sink function, we obtain the Green's function for the presence of two sink functions and it is given by
\begin{equation}
  \mathscr{G}_{2}(x,s+k_{r}|x_{0})=\mathscr{G}_{1}(x,s+k_{r}|x_{0})-\frac{k_{0}k_{2}\mathscr{G}_{1}(x,s+k_{r}|x_{2})\mathscr{G}_{1}(x_{2},s+k_{r}|x_{0})}{1+k_{0}k_{2}\mathscr{G}_{1}(x_{2},s+k_{r}|x_{2})}.  
\end{equation}
Obviously $\mathscr{G}_{1}(x,s+k_{r}|x_{0})$ will be calculated from Eq. (6).
So the above derivations suggest that by knowing $\mathscr{G}_{0}(x,s+k_{r}|x_{0})$ one can find the solution for any number of $\delta$ function sink presence on the potential energy curve analytically.
The general form of the Green's function solution for $n$ number of $\delta$ function sink will be 
\begin{equation}
  \mathscr{G}_{n}(x,s+k_{r}|x_{0})=\mathscr{G}_{n-1}(x,s+k_{r}|x_{0})-\frac{k_{0}k_{i}\mathscr{G}_{n-1}(x,s+k_{r}|x_{n})\mathscr{G}_{n-1}(x_{n},s+k_{r}|x_{0})}{1+k_{0}k_{n}\mathscr{G}_{n-1}(x_{n},s+k_{r}|x_{n})}. 
\end{equation}
Here we see that if we are able to find $\mathscr{G}_{0}$ for a potential energy curve then we can find the solution when there are $n$ numbers of $\delta$ sink function on PEC analytically.
\section{Calculation of Rate Constant}
In the following we provide a general procedure for calculating the average rate constant when multiple sink is presented on PEC. 
The average rate constant $K_{I}$ in Laplace domain has form ${\mathscr{P}_{e}}^{-1}(s=0)$ and it can be found from $\mathscr{P}_{e}(x,s)$ through the expression $\mathscr{P}_{e}(s)=\int_{-\infty}^{\infty} \mathscr{P}(x,s)dx $. Thus
$$K_{I}^{-1}=\mathscr{P}_{e}(0)$$
 For a potential energy curve $U(x)$ with initial distribution $P(x,0)=\delta(x-x_{0})$ we may find  
 \begin{equation}
     \mathscr{P}_{0}(s)=\mathscr{G}_{0}(s+k_{r}|x_{0})=\int_{-\infty}^{\infty} \mathscr{G}_{0}(x,s+k_{r}|x_{0})dx=\frac{1}{s+k_{r}}.
 \end{equation}
Which correspondingly refers that the total probability in the time domain is unity when there is no sink present on PEC. The Eq. (11) is also true for any value of $x_{0}$.
Therefor for one $\delta$ function sink at $x_{1}$ we get
\begin{equation}
    \mathscr{P}_{1}(s)=\mathscr{G}_{1}(s+k_{r}|x_{0})=\mathscr{G}_{0}(s+k_{r}|x_{0})-\frac{k_{1}\mathscr{G}_{0}(s+k_{r}|x_{1})\mathscr{G}_{0}(x_{1},s+k_{r}|x_{0})}{1+k_{1}\mathscr{G}_{0}(x_{1},s|x_{1})}
\end{equation}
Hence the inverse of average rate constant is
\begin{equation}
    K_{I1}^{-1} =\frac{1}{k_{r}}[1-\frac{k_{1}\mathscr{G}_{0}(x_{1},k_{r}|x_{0})}{1+k_{1}\mathscr{G}_{0}(x_{1},k_{r}|x_{1})}].
\end{equation}
$K_{I1}$ depends upon initial distribution $P(x,0)$. By using Eq. (9) we can calculate 
\begin{equation}
     K_{I2}^{-1}= K_{I1}^{-1}-\frac{k_{2}\mathscr{G}_{1}(x_{2},s+k_{r}|x_{0})}{1+k_{2}\mathscr{G}_{1}(x_{2},s+k_{r}|x_{2})}|_{s=0}K_{I1}^{-1}|_{x_{0}=x_{2}}.
\end{equation}
Here $K_{I1}^{-1}|_{x_{0}=x_{2}}=\int_{-\infty}^{\infty}\mathscr{G}_{1}(x,k_{r}|x_{2}) dx $. So in the above equation we are able to express $K_{I2}$ in terms of $K_{I1}$ and $\mathscr{G}_{1}(x,s+k_{r}|x_{0})$. The general expression for the rate constant in case of $n$ number of $\delta$ sink function can be written as
\begin{equation}
     K_{In}^{-1}= K_{In-1}^{-1}-\frac{k_{n}\mathscr{G}_{n-1}(x_{n},s+k_{r}|x_{0})}{1+k_{n}\mathscr{G}_{n-1}(x_{n},s+k_{r}|x_{n})}|_{s=0}K_{In-1}^{-1}|_{x_{0}=x_{n}}.
\end{equation}
However for nonzero of $k_{r}$ there is no problem as such in the evaluation of numerical value for the decay rate. But in the absence of radiative decay, i.e., $k_{r}=0$, divergences come in the evaluation of ${K_{I1}}^{-1}$ in Eq. (13) in the limit $s\rightarrow 0$ if the stationary distribution is nonzero. So in order to calculate the non-radiative decay rate, a special technique is required to overcome the divergence. For that we have to find $K_{I1}^{-1}$ in the limit $k_{r}\rightarrow0$.

\subsection{Flat Potential}
As a first example, we consider a simple case where a particle is influenced by a constant potential, for which the Green's function \begin{equation}
    \mathscr{G}_{0}(x,s+k_{r}|x_{0})=\frac{e^{-\sqrt{\frac{s+k_{r}}{D}}|x-x_{0}|}}{2\sqrt{D(s+k_{r})}}.
\end{equation}
Therefor the explicit form of $ \mathscr{P}_{1}(s)$ is
\begin{equation}
     \mathscr{P}_{1}(s)=\frac{1}{s+k_{r}}[1-\frac{k_{1}e^{-\sqrt{\frac{s+k_{r}}{D}}(x_{1}-x_{0})}}{2\sqrt{D(s+k_{r})}+k_{1}}].
\end{equation}
Hence average rate constant $K_{I1}$ for single sink is  
\begin{equation}
    {K_{I1}}^{-1}=\frac{1}{k_{r}}[1-\frac{k_{1}e^{-\sqrt{\frac{k_{r}}{D}}(x_{1}-x_{0})}}{2\sqrt{Dk_{r}}+k_{1}}].
\end{equation}
By using the above equation we can calculate $K_{I}^{-1}$ for more than one sink on the PEC. Numerical value of average rate constant have been calculated for different strength of sink function by using the below expression 
\begin{equation}
     {K_{In}}^{-1}= {K_{In-1}}^{-1}-\frac{k_{0}k_{i}\mathscr{G}_{n-1}(x_{n},s+k_{r}|x_{0})}{1+k_{0}k_{n}\mathscr{G}_{n-1}(x_{n},s+k_{r}|x_{n})}|_{s=0}K_{In-1}^{-1}|_{x_{0}=x_{n}}.
\end{equation}
In Figure 1, we plot $K_{I}$ against a number of the sink, $n$, on PEC for three values of $k_{i}$, viz. 1, 2 and 3. 
\begin{figure}[hbt!]
\centering
\begin{minipage}{.5\textwidth}
\includegraphics[height=4cm,width=6cm]{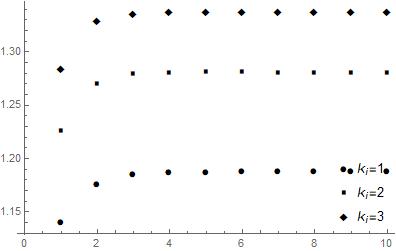}
    \caption{\small{\emph{ $K_{In}$ against number of sink function n. Assumed values are $D=1$, $k_{r}=1$ and $x_{i}=i$; i=1,2,...10, sink positions.}}}
    \label{kfenslk}
    \end{minipage}
\end{figure}
\newpage
This figure is showing $K_{I}$ gets saturated for the presence of a higher number of sink function. Also overall $K_{I}$ is increasing with increase of strength of each $\delta$ function sink. In Table 1, a small comparison between our method and Samanta-Ghosh's method have been made by calculating numerical values of $K_{I}$. 
 
\begin{table}[h!]
  \centering
\begin{tabular}{|c|c|c|}
    \hline
    n & $ K_{IS}$ & $K_{IO}$  \\
    \hline
  1 & 1.13977 & 1.13977 \\
  \hline
   2 & 1.17566 &  1.17566 \\
  \hline
   3 & 1.18468 &  1.18462  \\
  \hline
   4 &  1.18693  & 1.18682 \\
  \hline
   5 & 1.18749  &  1.18737 \\
  \hline
   6 & 1.18763 & 1.18750\\
  \hline
  
\end{tabular}
 \caption{\small{\emph{A comparison of rate constant with number of sink function n under the same value of rest of parameters. $K_{IS}$ is from Samanta-Ghosh's method and $K_{IO}$ is from our method.}}} \label{tab:sometab}
\end{table}
\subsection{Linear Potential}
We can get the Green's function for the linear potential $U(x)=|x|$ from the following equation
\begin{equation}
     [s+k_{r}-\mathscr{L}+S(x)]\mathscr{P}(x,s)=\delta (x-x_{0}).
\end{equation}
with $\mathscr{L}=D\frac{d^2}{d x^2}+ \frac{D}{k_{B}T}\frac{d}{d x}\frac{|x|}{x}$. By using proper boundary condition, the Green's function for the linear or V shaped potential will be \citep{chase2016analysis}
\begin{equation}
    \mathscr{G}_{0}(x,s+k_{r}|x_{0})=\frac{e^{-(|x|-|x_{0}|)/2l}}{\Gamma \sqrt{1+\frac{4ls+k_{r}}{\Gamma}}}[e^{-\sqrt{1+\frac{4ls+k_{r}}{\Gamma}}|x-x_{0}|/2l}+ \frac{e^{-\sqrt{1+\frac{4ls+k_{r}}{\Gamma}}(|x|+|x_{0}|)/2l}}{\sqrt{1+\frac{4ls+k_{r}}{\Gamma}}-1}].
\end{equation}
Where $\Gamma=D/k_{B}T$ and $l=D/\Gamma$. The solution for generalized sink can be determined by using Eq. (10). For non radiative decay rate calculation the expression for, in case of single $\delta$ function sink, average rate constant from Eq. (13)
\begin{equation}
    K_{I1}^{-1} = \lim_{s\rightarrow0}\frac{1}{s}[1-\frac{k_{1}\mathscr{G}_{0}(x_{1},s|x_{0})}{1+k_{1}\mathscr{G}_{0}(x_{1},s|x_{1})}].
\end{equation}
By using the Eq. (21), with $k_{r}=0$, and after calculating the limiting value of right hand site of of the above equation we find
\begin{equation}
    K_{I1}^{-1} = \frac{1}{k_{1} \Gamma}(2lk_{1} e^\frac{|x_{1}|}{l}-2lk_{1} e^\frac{|x_{0}|+|x_{1}|-|x_{0}-x_{1}|}{2l}+2 l \Gamma e^\frac{|x_{1}|}{l}+k_{1}|x_{0}|- k_{1}|x_{1}|).
\end{equation}
We also calculate $\mathscr{G}_{1}(x,s+k_{r}|x_{j})$ in the limit $s\rightarrow0$, which will be needed to calculate rate constant for a higher number of sink. The relevant expression is 
\begin{equation} \label{eq1}
\begin{split}
 \lim_{s\rightarrow0}\mathscr{G}_{1}(x,s+k_{r}|x_{j}) & =\frac{1}{\Gamma} e^-\frac{2|x|+|x-x_{j}|+|x-x_{1}|+|x_{j}-x_{1}|}{2l} \\
 &\times [e^-\frac{|x|+|x_{j}|+|x-x_{1}|+|x_{j}-x_{1}|}{2l}\\
 &  - e^-\frac{|x-x_{j}|+|x_{j}|+|x-x_{1}|+|x_{1}|}{2l}\\
 &  - e^-\frac{|x|+|x-x_{j}|+|x_{j}-x_{1}|+|x_{1}|}{2l}\\
 & +(1+\frac{\Gamma}{k_{1}})e^-\frac{|x-x_{j}|+|x-x_{1}|+|x_{j}-x_{1}|+2|x_{1}|}{2l}].
\end{split}
\end{equation}
The above expression is derived by employing the Green's function of the linear potential. So the form of inverse of nonradiative decay rate for two $\delta$ function sink at positions $x_{1}$ and $x_{2}$ is nothing but the Eq. (14) with $k_{r}\rightarrow0$. In that equation we use Eq. (24) to calculate the numerical value of $K_{I2}$. So we can easily calculate the rate constant for a higher number of $\delta$ function sink by using our recursion formula in Eq. (15) and the Eq. (24). In Figure (2), we have plotted $K_{In}$ against a number of sinks $n$ on PEC for the cases, one for when initial distribution is at the left side of all sinks and another for when it is at the right side of all sinks.  
\begin{figure}
  \begin{subfigure}[b]{0.4\textwidth}
    \includegraphics[width=\textwidth]{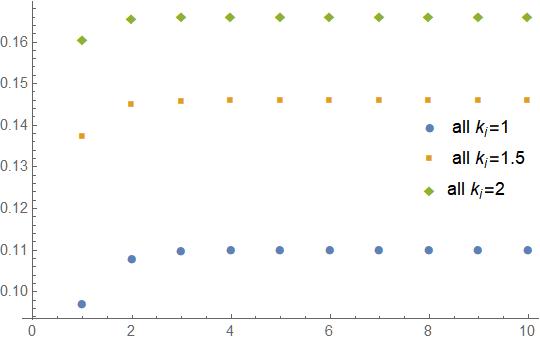}
    \caption{}
    \label{fig:1}
  \end{subfigure}
  \begin{subfigure}[b]{0.4\textwidth}
    \includegraphics[width=\textwidth]{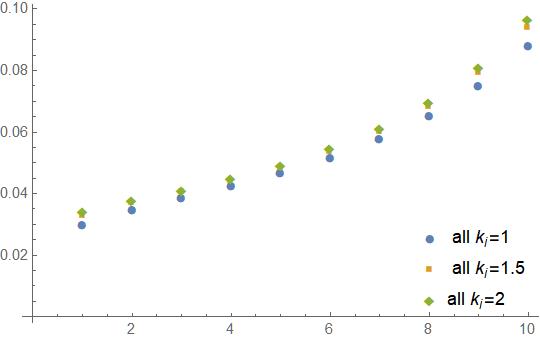}
    \caption{}
    \label{}
  \end{subfigure}
  \caption{\small{\emph{ $K_{In}$ against n and initial distribution is at the right side of all sinks. Assumed values are $D=0.5$, $l=1$ and $x_{i}=i$; i=1,2,...10, sink positions. (a) When P(x,0) is placed at the origin of PEC, i.e, $x_{0}=0$. (b) P(x,0) is positioned right side of all sinks. We take $x_{0}=15$.}}}.  
\end{figure}
\subsection{Parabolic Potential}
In the following, we get the propagator for the parabolic potential $U(x)=\frac{1}{2}\mu {\omega}^2 x^2$, $\mu$ is mass and $\omega$ is frequency, by solving the following equation: 
\begin{equation}
    [s+k_{r}-D\frac{\partial^2}{\partial x^2}-\frac{D}{K_{B}T}\frac{\partial}{\partial x}(\frac{dU(x)}{dx})]\mathscr{G}_{0}= \delta(x-x_{0})
\end{equation}
to get
\begin{equation}
     \mathscr{G}_{0}(x,s+k_{r}|x_{0})= F(z,s+k_{r}|z_{0})/(s+k_{r}) ,
\end{equation}
with
\begin{equation}
     F(z,s+k_{r}|z_{0})=D_{\nu}(z)D_{\nu}(-z_{0})exp[({z_{0}}^2-{z}^2)/4]\Gamma(1-\nu)[B/(2\pi D)]^{1/2}.
\end{equation}
In the above we have introduced new variable $z$ such that $z=x\sqrt{\frac{B}{D}}$ and $z_{i}=x_{i}\sqrt{\frac{B}{D}}$ \citep{sebastian1992theory}. We also have consider $B=\mu\omega^2 D/(k_{B}T)$ and $\nu=-s/B$. $D_{\nu}$ are parabolic cylinder functions and $\Gamma(\nu)$ is the gamma function.
\begin{figure}[hbt!]
\centering
\begin{minipage}{.5\textwidth}
\includegraphics[height=4cm,width=6cm]{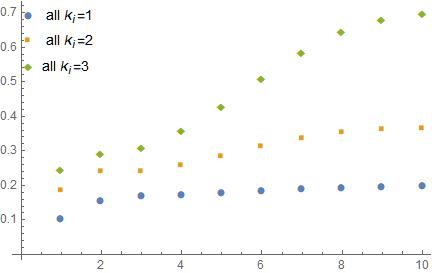}
    \caption{\small{\emph{ $K_{In}$ against number of sink function n. Assumed values are $D=1$, $x_{i}=i$; i=1, 2, 3...10. and $B=0.1$. Initial distribution at $x_{0}=0$.}}}
    \label{kfenslk}
    \end{minipage}
\end{figure}
Now we consider Eq. (13) for this PEC, which is
\begin{equation}
    K_{I1}^{-1}= \frac{1}{k_{r}}[1-\frac{k_{1}F(z_{1},k_{r}|z_{0})}{k_{r}+k_{1}F(z_{1},k_{r}|z_{1}) }].
 \end{equation}
 
In the limit $\nu \rightarrow0$ we know $D_{\nu}(z)=0$. So in this limit of $\nu$, as $k_{r}\rightarrow 0$, we find   $F(z_{1},k_{r}|z_{0})$ and $F(z_{1},k_{r}|z_{1})\rightarrow exp(-{z_{1}}^{2}/2)[B/(2 \pi D)^{1/2}]$ so that $k_{1}F(z_{1},k_{r}|z_{0})/k_{r}+k_{1}F(z_{1},k_{r}|z_{1})$ $  \rightarrow 1$. Therefor Eq. (28) can be written as
\begin{equation}
     K_{I1}^{-1}=\lim_{k_{r}\rightarrow 0}[(\frac{k_{0}k_{1}F(z_{1},k_{r}|z_{0})}{k_{r}+k_{0}k_{1}F(z_{1},k_{r}|z_{1})})_{k_{r}=0}-\frac{k_{0}k_{1}F(z_{1},k_{r}|z_{0})}{k_{r}+k_{0}k_{1}F(z_{1},k_{r}|z_{1})}]
\end{equation}
We consider that $x_{0} < x_{1}$, without loss of generality, so that the initial distribution is placed to the left of the sink and the above equation can farther be evaluated as 
\begin{equation}
    K_{I1}^{-1}=\sqrt{\frac{2\pi D}{B}} \frac{e^{{z_{1}}^{2}/2}}{k_{1}}+\frac{exp(\frac{{z_{0}^{2}-z_{1}^{2}}}{4})}{B}[\frac{d}{d\nu}(\frac{D_{\nu}(-z_{0})}{D_{\nu}(-z_{1})})]_{\nu=0}.
\end{equation}
After using proper expression for $D_{\nu}(-z)$ we can obtain
\begin{equation}
    K_{I1}^{-1}=\sqrt{\frac{2\pi D}{B}} \frac{e^{{z_{1}}^{2}/2}}{k_{1}}+\frac{1}{B}\sqrt{\frac{\pi}{2}}\int_{z_{0}}^{z_{1}}exp(z^{2}/2)(1+ erf(z/\sqrt{2}))dz.
\end{equation}
Where $erf(z)$ is a error function and has a form, $erf(z)=\frac{2}{\sqrt{\pi}}\int_{0}^{z}e^{-t^{2}} dt$. We can also find
\begin{equation}
    \lim_{s\rightarrow0}\mathscr{G}_{1}(x,s|x_{j})= \frac{e^{-\frac{\sqrt{B/D}}{2}(x^2-x_{j}^{2})}}{k_{1}}.
\end{equation}
Using the above general expression and the Eq. (31) we may calculate $K_{I2}$ and the rate constant corresponding to a higher number of $\delta$ function sink when $k_{r}=0$. We have calculated nonradiative decay rate numerically, graphically shown in Figure (3), in presence of more than one sink on the PEC when $P(x,0)$ is sharply picked at the center of the PEC and sinks are considered on the right side of $P(x,0)$. 
\section{ Summary and Conclusions}
Our Eq. (10) is quit general for generalized sinks present on PEC in Laplace domain. Since the form of sink is most likely to be a Gaussian function and we can approximate this form by choosing strength of each $\delta$ function sink properly. Eq. (15) represents the rate constant for a generalized sink. If we are able to find the nonradiative decay rate for a localized sink, then It is very easy to calculate a nonradiative decay rate for a generalized sink by using our method than the Samanta and Ghosh's method.     
\section{ Acknowledgments}
The author (C.S.) would like to thank Indian Institute of Technology Mandi for Half-Time Research Assistantship fellowship. The author is grateful to Aniruddha Chakraborty for suggestion and valuable discussion. \\




\refname


\begin{thebibliography}{00}
\bibitem{wilemski1974diffusion}
G. Wilemski and M. Fixman, J. Chem. Phys. 60(1974)866.
\bibitem{wilemski1974diffusion2}
G. Wilemski and M. Fixman, J. Chem. Phys. 60(1974)878.
\bibitem{ovchinnikova1982frequency}
M. Ya. Ovchinnikova, Theor. Exp. Chem. 17(1982)507.
\bibitem{zusman1983theory}
L. D. Zusman, Chem. Phys. 80(1983)29.
\bibitem{sumi1986dynamical}
H. Sumi and R. A. Marcus, J. Chem. Phys. 84(1986)4894.
\bibitem{sumi1995}
H. Sumi, J. Mol. Liq. 65/66(1995)65.
\bibitem{bagchi1983theory}
B. Bagchi, G.R. Fleming, D.W. Oxtoby, J. Phys. Chem. 78(1983)7375.
\bibitem{bagchi1990dynamics}
B. Bagchi, G.R. Fleming, J. Phys. Chem. 94(1990)9.
\bibitem{sebastian1992theory}
K.L. Sebastian, Phys. Rev. A. 46(1992)R1732.
\bibitem{chakravarti1993theory}
N. Chakravarti, K.L. Sebastian, Chem. Phys. Lett. 204(1993)496.
\bibitem{samanta1990deuterium}
A. Samanta, S.K. Ghosh, H.K. Sadhukhan, Chem. Phys. Lett. 168(1990)410.
\bibitem{wade1998electrostatic}
R.C. Wade, R.R. Gabdoulline, S.K. L{\"u}demann, V. Lounnas, Proc. Natl. Acad. Sci. 95(1998)5942. 
\bibitem{livesay2003conservation}
D.R. Livesay, P. Jambeck, A. Rojnuckarin, S. Subramaniam, Biochemistry. 42(2003)3464.
\bibitem{giuggioli2006theory}
L. Giuggioli, G. Abramson, V.M. Kenkre, R.R. Parmenter, T.L. Yates, J. Theor. Biol. 240(2006)126.
\bibitem{szabo1984localized}
A. Szabo, G. Lamm, G.H. Weiss, J. Stat. Phys. 34(1984)225.
\bibitem{samanta1993exact}
A. Samanta, S.K. Ghosh, Phys. Rev. E. 47(1993)4568.
\bibitem{chase2016analysis}
M. Chase, K. Spendier, VM Kenkre, J. Phys. Chem. B. 120(2016)3072.
\end{thebibliography}
\end{document}